\documentclass[journal]{IEEEtran}

\usepackage[utf8]{inputenc}
\usepackage{graphicx}
\usepackage{url}
\usepackage{footnote}
\usepackage{booktabs}
\usepackage{amsmath} % Required for \varPsi below
\usepackage{amsfonts}
\usepackage{multirow}

\begin{document}
\title{Architectural exploration \\of heterogeneous memory systems}
%[OK]GABRIEL: yo en principio dejaría el título del abstract que enviamos
%using\\
%gem5 framework}

\author{Marcos~Horro,
        Gabriel~Rodríguez,
        Juan~Touriño,
        Mahmut~T.~Kandemir % <-this % stops a space
\thanks{M. Horro, G. Rodríguez and J. Touriño are with the Department
of Electronics and Systems, Universidade da Coruña,
\texttt{\{marcos.horro,grodriguez,juan\}@udc.es}.}% <-this % stops a space
\thanks{M. T. Kandemir is with the Department of Computer Science and Engineering, Pennsylvania State University,
\texttt{kandemir@cse.psu.edu}.}}

\maketitle

\begin{abstract}
%The future of computer architecture presents important challenges. Scaling is reaching the physic limits of transistors and then it is %crucial to develop new alternatives to mitigate this stagnation. Heterogeneous systems are a feasible solution. gem5 framework provides %several architectures to simulate and also allows to configure any memory hierarchy. Thus, we have adapted and integrated scratchpad %memories onto the simulator and, combined with a linear programming solver to decide on which memory module must be allocated a certain %variable in a program, we have obtained improvements in the energy consumption regarding architectures without scratchpad memories.

%[OK]GABRIEL: he reescrito esto. Échale un vistazo.
Heterogeneous systems appear as a viable design alternative for the dark silicon era. In this paradigm, a processor chip includes several different technological alternatives for implementing a certain logical block (e.g., core, on-chip memories) which cannot be used at the same time due to power constraints. The programmer and compiler are then responsible for selecting which of the alternatives should be used for maximizing performance and/or energy efficiency for a given application. This paper presents an initial approach for the exploration of different technological alternatives for the implementation of on-chip memories. It hinges on a linear programming-based model for theoretically comparing the performance offered by the available alternatives, namely SRAM and STT-RAM scratchpads or caches. Experimental results using a cycle-accurate simulation tool confirm that this is a viable model for implementation into production compilers.
\end{abstract}


% \fi
\IEEEpeerreviewmaketitle

\section{Introduction}\label{sec:intro}

%TODO:
%1.- FIX FIGURES RESULTS
%2.- EXPLAIN CORRELATION
%3.- ?

\IEEEPARstart{T}{he} end of Dennard scaling has brought processor performance to a near standstill in recent years, pushing architects towards designing increasingly parallel computers. Even with this paradigm shift, the power wall is expected to soon limit multicore scaling. A dark silicon future~\cite{bib:dennard} has been proposed, in which chips will incorporate a high number of very specialized hardware such as vector execution units, tunable memory hierarchies and even different heterogeneous core technologies. In this paradigm, users, compilers and runtime 
%[OK]GABRIEL
%have to 
cooperate to choose which parts of the core to use, under a certain power budget, for executing a given application under given QoS restrictions.
%[OK]GABRIEL: elimino esta referencia que ya aparecía en la frase anterior
%~\cite{bib:dennard}.

This paper focuses on the analysis and simulation of heterogeneous memory hierarchies using the gem5 simulator, a widely used tool developed by key players in the industry such as ARM, Intel, or Google. Its biggest potential is the ability to create new components or modify existing ones in order to perform architectural exploration. Extensions to simulate scratchpad memories in this framework are proposed, and used to assess the power-performance trade-off of different memory configurations and technologies. Profiling data is fed to an operational research framework that decides which memory units to enable from an available pool. This mathematical framework takes variable placement decisions by checking whether an access presents reuse that is easily exploitable by a regular cache, or whether the LRU algorithm and cache conflicts make it advisable to manually allocate the variable to a scratchpad to better take advantage of the available locality. Taking into account the hardware characteristics of the different available memory modules, access and transfer costs for each possible allocation are calculated, and a final allocation is decided considering also memory sizes and power budgets. 

%[DONE]GABRIEL: para el último párrafo de la introducción se sigue siempre un formato bastante estándar: se enumeran las siguientes secciones del artículo y se resumen en una frase. Échale un vistazo a cualquier paper en mi web y copia la idea. La frase que tienes actualmente aprovéchala para el resumen de la sección de experimentales. Ojo, usa referencias (\ref{sec:state-of-the-art}) y etiquetas (\label{sec:state-of-the-art}) para garantizar la consistencia.
This paper is organized as follows. Section~\ref{sec:scratch} introduces scratchpad memories. Section~\ref{sec:gem5} analyzes the gem5 framework and discusses  the simulation of scratchpad memories in the system. The mathematical model for data allocation is presented in Section~\ref{sec:math}. Experimental results using SRAM caches and STT-RAM scratchpad memories demonstrate the feasibility of the proposed approach, as shown in Section~\ref{sec:results}. Finally, Section~\ref{sec:discussion} concludes the paper.

%[OK]GABRIEL: voy a cambiar el título de esta sección tentativamente
%\section{State of the art}
\section{Scratchpad memories}\label{sec:scratch}
%[DONE]GABRIEL: cambiar el principio de este párrafo para que no sea tan abrupto. La idea es presentar las memorias scratchpad para el lector que no sepa de qué se trata.
The importance and complexity of the on-chip memory hierarchy has increased with the advances in processor performance~\cite{bib:yan}, in an attempt to bridge the gap between memory and processor speeds. Nowadays it is possible to find several different types of memories integrated in a single hierarchy, e.g., private vs. shared ones, or exclusive vs. inclusive. Scratchpad memories (SPMs) are one type of fast, random-access memories which are sometimes used as an alternative to cache memories.
The main feature of scratchpads is its programmability, i.e., the possibility of handling the data allocated using special-purpose instructions placed by the compiler or programmer. In contrast, the contents of cache memories are controlled automatically by the hardware, for instance using the LRU algorithm. The use of scratchpad memories is widely extended in embedded systems, e.g. real time systems.
%[DONE]GABRIEL: no me acaba de convencer esta explicación. la has leído en alguna parte?
%where the use of cache memories is prohibitive since the data flow does not favor the locality and even causing the use of cache memories to be unprofitable in some cases.
%MARCOS: la idea venía del siguiente párrafo al que adjunto cita
SPM guarantees a fixed access latency whereas an access to the cache may result in a miss thereby incurring longer latency due to off-chip access~\cite{spmvscache}~\cite{spmvscache2}. This unpredictability of caches is undesirable to meet hard timing constraints. Besides, scratchpad memories, having no tag array, are potentially more efficient energy- and performance-wise. Nevertheless, cache memories are largely used in general purpose processors, mainly because efficiently employing a scratchpad memory implies recompiling the code for different SPM features (e.g., size), while the LRU algorithm employed by caches is capable of automatically exploiting locality in a reasonable way.  The use of one or the other flavor of on-chip memory becomes then a performance/effort choice. To exploit this trade-off,  several specific purpose architectures have implemented scratchpad memories:

				%GABRIEL: se podría añadir una cita a la Playstation 3 y a las TVs de Toshiba?
                %[MARCOS]: añadidas citas
\begin{itemize}
	\item Cell IBM~\cite{bib:cell}: this architecture was used in the nodes of the MareNostrum supercomputer at the Barcelona Supercomputing Center~\cite{bsc} and in the main core of PlayStation 3, as well as in premium Toshiba TVs~\cite{bib:cellexamples}, amongst others.%[OK]GABRIEL: añade una cita al BSC
    \item PlayStation 2~\cite{bib:ps2}: this console included small scratchpad memories managed by the CPU and GPU.
    \item Digital Signal Processor (DSP)~\cite{bib:dsp}: SODA's architecture includes a private scratchpad memory for each processing unit and a global scratchpad memory that can be accessed for any unit.
    \item Knights Landing's architecture Intel Xeon Phi~\cite{bib:knights}: this architecture uses MCDRAM memories. The processor has access to an addressable memory with high bandwidth, and therefore the concept is similar to that of scratchpad memories. However it is also remarkable that, in contrast with the previous examples, Intel Xeon Phi is a manycore architecture.
\end{itemize}

There has also been at least an implementation of scratchpad memories in a general purpose architecture, as Cyrix 6X86MX~\cite{ibm} already implemented a ``scratchpad RAM''. This 32-bit x86-compatible microprocessor, released in 1996,
%with a superpipelined architecture,
implemented a large Primary Cache of 64kB. This cache could be turned into a scratchpad RAM memory. The cache area set aside as scratchpad memory acted as a private memory for the CPU and did not participate in cache operations. 
%Then, the concept we have is very similar, but instead turning a cache into it, having a dedicated area on the die for it. %[DONE]GABRIEL: puedes dar un par de detalles de este SPM? (tamaño, tecnología, cualquier cosa para que este párrafo quede un poco más completo).

%[DONE]GABRIEL: creo que cogería el resto de la sección y lo movería a la siguiente, para dejar todo el contenido de gem5 e implementación de scratchpads en el mismo sitio.

\section{Integration of scratchpad memories in gem5}\label{sec:gem5}
%On the other hand, 
Simulators are crucial tools in architectural development and exploration,  since they allow reasonably accurate estimations with almost negligible cost in comparison with lithographic or FPGA solutions. Thus, there have appeared many approaches with different granularity focused on concrete subsystems and other with holistic nature. Table~\ref{table:simcomparison} compares different simulators with both private licenses, as Simics, and open source licenses, which are the rest. 

\begin{table}[hbt]
	\normalsize
	\centering
    \caption[Caption]{Differences between simulators according to the following criteria: whether the platform is or not in current development, Instruction Set Architecture (ISA) supported and accuracy. Data extracted from~\cite{gem5accuracy}~\cite{gem5:errors}}
\begin{tabular*}{8cm}{p{1.3cm}p{0.5cm}p{3.2cm}p{2.7cm}}
		\toprule
		\textbf{Sim.}    & \textbf{Dev.} & \textbf{ISA(s)} & \textbf{Accuracy}       \\
		\midrule
        \midrule
		Simics       & Yes         & Alpha, ARM, M68k, MIPS, PowerPC, SPARC, x86 & Functional     \\ \midrule
		SimFlex      & No          & SPARC, x86 (requires Simics)              & Cycle \\ \midrule
		GEMS         & No          & SPARC (requires Simics)                   & Timing         \\ \midrule
		m5			 & No		   & Alpha, MIPS, SPARC				   	      & Cycle \\ \midrule
		MARSS        & No         & x86 (requires QEMU)                       & Cycle \\ \midrule
		OVPsim       & Yes         & ARM, MIPS, x86                            & Functional     \\ \midrule
		PTLsim       & No          & x86 (requires Xen and KVM/QEMU)          & Cycle \\ \midrule
		Simple
Scalar & No          & Alpha, ARM, PowerPC, x86                  & Cycle\\ \midrule
		gem5         & Yes         & Alpha, ARM, MIPS, PowerPC, SPARC, x86     & Cycle \\\bottomrule
	\end{tabular*}
	\label{table:simcomparison}
\end{table}

gem5 is a framework for performing cycle-accurate computer architecture simulations. The main reasons for choosing gem5 over the rest in this work have been: (i) its current development and implication of different organizations such as Google, Intel or ARM, as well as its large community of users; (ii) its open source license; and (iii) the possibility of simulating different ISAs. In addition, the modularity of the platform allows an easy modification of the system and the integration of new components. It provides two modes of operation: full system and system call emulation mode. The latter is interesting for executing benchmarks without loading an OS image. Regarding memory hierarchy simulation, there are two modes: classic and Ruby. An advantage of the classic mode is its simplicity. In the Ruby system memories are specified as finite state machines, and it is focused on studying the impact of using different cache coherence protocols.

In this  paper, the main goal of including scratchpad memories in the system is to reduce energy consumption. Cache memory energy constitutes a major part of modern processor consumption~\cite{bib:cacheenergy}~\cite{gabri}, the main reason being the need to activate both the tag and data regions to obtain a certain line, while a random-access memory only needs a single array access. Besides, cache misses also imply penalties in energy consumption.

Moreover, the concept of scratchpad memories does not differ from conventional memories, i.e., RAM memory, as their contents can be allocated and programmed by the application code. Thus, in order to implement scratchpad memories (SPM) in gem5, the starting point has been the memory modules already implemented (\texttt{SimpleMemory} class). We have adapted this implementation in order to add new parameters, such as latencies.

\begin{figure}[htb]
	\begin{center}    	
    	\includegraphics[scale=0.5]{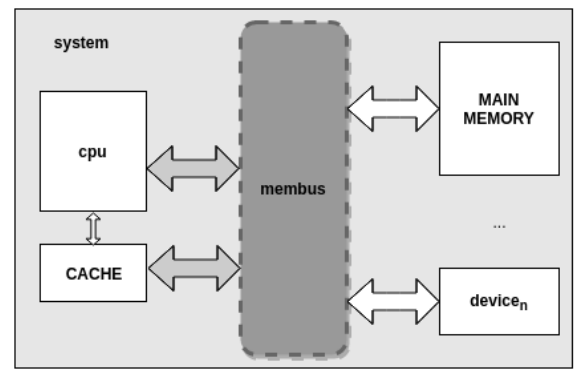}
    \end{center}
    \caption{Original architectural scheme in gem5}
    \label{fig:archconvgem5}
\end{figure}

\begin{figure}[h]
	\begin{center}
    	\includegraphics[scale=0.4]{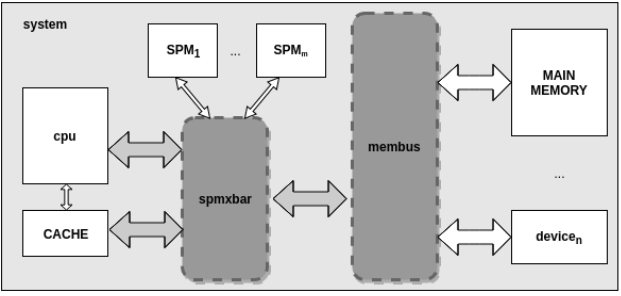}
    \end{center}
    \caption{Architectural scheme including SPM in gem5}
    \label{fig:archspmgem5}
\end{figure}

%GABRIEL: esta línea queda huérfana al final de la página 2
Regarding the connection of these memories to the system, we considered enabling a new and specific port in the CPU. However, this was deemed a nonportable approach, since each different processor included in gem5 would need to be independently
%GABRIEL: técnicamente, no podríamos modificar únicamente BaseCPU? No lo pregunto para cambiar el paper sino pensando en trabajo futuro.
adapted. For this reason, we decided to use the already implemented crossbars. In the gem5 classic memory system, CPUs are provided with three ports: a system port and two ports for data and instruction caches. The system port, by default, is connected to the main bus (\texttt{membus}) where all memory devices are connected, e.g. main memory, I/O devices, etc. We modified these connections, creating a new crossbar which induces no latencies nor overheads where scratchpad memories are connected. This configuration connects CPU, cache memories, \texttt{membus} and scratchpad memories. Basically, this crossbar acts like a ``bridge'' between all the components.
%and 
%GABRIEL: esta frase queda un poco repetitiva, pero no estoy seguro de cómo arreglarla
%the main bus used before to connect the main bus
%MARCOS: así?
Figures \ref{fig:archconvgem5} and \ref{fig:archspmgem5} illustrate the difference between the original configuration and the modified architecture proposed.

The last step in order to integrate these memories in the system is the possibility to access their content. For this purpose the original ISA has been modified (in our case we chose x86 for familiarity reasons), adding an instruction to explicitly allocate a range of physical addresses onto the scratchpad memory. In order to simplify this allocation, the physical range chosen 
%corresponds
%GABRIEL: esta frase es confusa
%to the next address of main memory until the addition of it and the size of the scratchpad memory given.
starts after the main memory range. This way, in order to integrate more than one scratchpad memory in the system, the next SPM's range starts right after the scratchpad memory instantiated before, i.e:

\begin{equation*}
\begin{aligned}
	\text{range}(SPM_{1}) = & [|\text{main memory}|, |\text{main memory}| + |SPM_1|]\\
    \text{range}(SPM_{i}) = & [\text{last\_addr}(SPM_{i-1})+1, \\
                            & \text{last\_addr}(SPM_{i-1}) + |SPM_{i}| + 1]\\
    \forall i > & 1\\
\end{aligned}
\end{equation*}

Thus, this instruction maps dynamically the region given, returning a reference in the program. This instruction has been generalized, allowing the reservation of memory in different scratchpad memories.

\section{Mathematical Model}\label{sec:math}

Given that the proposed architecture integrates programmable memories, it is necessary to decide where to allocate a specific variable of a program. In other words, decide whether a variable should reside into main memory and be accessed through a regular cache, or whether it should be copied to one of the available scratchpad memories.

We propose a linear programming system that minimizes a target function to control memory allocation. One advantage of this approach is the possibility to change the objective function easily to target performance and even add or remove restrictions to model total power limitations, memory sizes, or other types of QoS. In the scope of this paper, the target function is the dynamic energy ($E_{DYN})$, calculated as the addition of the energies of each individual access,
%[?]GABRIEL: qué es el "energy consumption factor of each memory"?
either read or write, multiplied by the energy consumption factor of each memory (i.e. the dynamic energy consumed by each read or write) and a binary decision variable. The introduced restrictions are: (i) a variable can only be allocated to a single memory, and (ii) the total size of the variables allocated to a scratchpad memory must be less or equal to the size of such memory. The size of each variable depends on the type of the variable (e.g. double, integer, etc.) and the number of elements in case of a k-dimensional vector.
%[OK]GABRIEL: el tamaño de las variables es irrelevante, no? Podrías considerar el tipo de datos en las ecuaciones. Por otra parte, los parámetros de tamaño usados en los tests es mejor discutirlos en la siguiente sección.
%We are considering variables of eight bytes, since we are working with double value, while the size of scratchpad memories begins in 1MB to 16MB, as we will discuss in the following section. 
This proposal is captured in the equations in Figure~\ref{eq:firstapproach}, where $|\psi|$ is the size of the variable $\psi$, $|SPM_i|$ is the size of $SPM_i$, $f_i(\psi)$ refers to the energetic cost of accessing scratchpad $i$ for variable $\psi$, and analogously $h(\psi)$ refers to the energetic cost of accessing main memory. $N_r$ and $N_w$ refer to the number of reads and writes respectively.

\begin{figure}[h]
  \begin{equation*}
      \begin{aligned}
      \psi \in \Psi & \equiv \text{ set of variables}\\
      \text{min } E_{DYN} & = \sum_{\psi\in\Psi}{\Big(M(\psi)h{(\psi)} + \sum_{i=1}^{n}{S_{i}(\psi)f_{i}(\psi)}\Big)}\\
      &S_{i}(\psi) = 
      \begin{cases}
      0 & \psi\text{ in }SPM_{i}\\
      1 & \psi\text{ not in }SPM_{i}
      \end{cases}\\
      &M(\psi) = 
      \begin{cases}
      0 & \psi\text{ in main memory}\\
      1 & \psi\text{ not in main memory}
      \end{cases}	\\  
      & M(\psi) + \sum_{i=1}^{n}{S_{i}(\psi)} = 1; \forall \psi \in \Psi\\
      & \sum_{\psi\in\Psi}{S_{i}(\psi)|\psi|} \leq |SPM_{i}|, \forall i \in [1,n]\\
      \text{where } & \\
      |\psi| = & \text{ type}(\psi)*N_{\psi}\\
      f_{i}(\psi) = & \sum_{\psi\in\Psi}{E_{SPM\_r}(i)N_{r}(\psi)+E_{SPM\_w}(i)N_{w}(\psi)}\\
      h(\psi) = & \sum_{\psi\in\Psi}{E_{MM\_r}N_{r}(\psi)+E_{MM\_w}N_{w}(\psi)}\\
      \end{aligned}
  \end{equation*}
  \caption{Linear programming system}
  \label{eq:firstapproach}
\end{figure}

%[OK]GABRIEL: añade aquí un pequeño párrafo explicando la notación de la Fig. 3 que no es explicada directamente en la figura, básicamente |\phi|, |SPM_i|, fi() y h().

Since energy savings from using scratchpad memories come from the reduction of accesses to main memory, the proposed system would recommend allocating the variables that fit into the scratchpad memory.
%[OK]GABRIEL
%Since we are comparing the dynamic energy that a scratchpad memory would consume against the main's memory, in principle if all the variables fit on the scratchpad memory, this system would always choose to allocate all variables in scratchpad memory, with the premise that main's memory energy is higher than scratchpad's. 
This can be detrimental in some cases, e.g., if an LRU cache
%GABRIEL
%a cache memory using LRU 
could take advantage efficiently of the locality present in the pattern of accesses of a given variable (cache-friendly access). Hence, we decided to introduce a new restriction: force a particular variable to be allocated to main memory if 
%GABRIEL
%satisfies the previous premise
the code accesses it in a cache-friendly way. This restriction is reflected in Equation \eqref{eq:cachef}.

\begin{equation}
	\label{eq:cachef}
	\begin{aligned}
		C(\psi) = &  
		\begin{cases}
		0 & \psi\text{ not \textit{cache-friendly}}\\
		1 & \psi\text{ \textit{cache-friendly}}
		\end{cases}	\\
		M(\psi) \geq &  \text{ }C(\psi) \hspace{0.2cm} \forall \psi \in \Psi \\
	\end{aligned}
\end{equation}

The value of this binary variable allows to introduce external information, e.g., data obtained by profiling or analyzing data at compile time.
%GABRIEL
%[OK]In our case, in order to avoid exhaustive profiling, we have decided to determined 
In our current system, the value of this variable is determined in a very simple analytical way: a regular access to a variable is considered cache-friendly if two consecutive accesses to a k-dimensional array ($k>1$) in a loop nest may reside in the same cache line. 

%[OK]GABRIEL: no me convence, nosotros sólo analizamos accesos afines, pero se podría analizar para cualquier tipo de accesos. Por otra parte, un acceso afín puede depender de N variables de lazo, por lo que la ecuación no es general. Como sólo probamos códigos afines, y además el sistema no es definitivo sino que lo hacemos a mano como solución temporal, no es necesario explayarse en este punto.
% For this purpose, a necessary condition but not sufficient if the affinity in the index, as it is explained in the Equation \eqref{eq:affinity}:
% \begin{equation}\label{eq:affinity}
% \begin{aligned}
% ai + b \hspace{1cm}& a,b \in \mathbb{R}\\
% \end{aligned}
% \end{equation}

As a corollary, the following expressions are exposed and commented briefly:

\begin{equation*}\label{eq:thirdapp}
\begin{aligned}
%[OK]A[i,j] \equiv & \text{ affine index and cache-friendly}\\
A[i,j] \equiv & \text{ regular, cache-friendly access}\\
%[OK]A[i, B[j]] \equiv & \text{ not an affine index therefore not cache-friendly}\\
A[i, B[j]] \equiv & \text{ irregular access, considered non cache-friendly}\\
%[OK]A[i, j*8] \equiv & \text{ affine but not necessarily cache-friendly}\\
A[i, j*8] \equiv & \text{ regular, but not necessarily cache-friendly access}\\
\end{aligned}
\end{equation*}

%[OK]It is remarkable this last example. It would not be cache-friendly since we are working with eight byte values (type double) and since the cache line is, commonly, 64 bytes. This, we can rewrite a cache-friendly access as:
In the last case the cache-friendliness of the access depends on the data type and cache line size. For instance, working with 8-byte double values and 64-byte cache lines two consecutive accesses would never reside in the same cache block. In the general case:

\begin{equation}\label{eq:thirdapp2}
	\begin{aligned}
		\exists a_{i}\hspace{0.2cm}/\hspace{0.2cm} & addr(a_{i+1}) - addr(a_{i}) \leq |cb|\\
	\end{aligned}
\end{equation}

Where $a_{i}$ and $a_{i+1}$ are two consecutive accesses, $addr(x)$ is the address of $x$ and $|cb|$ is the cache line size. With the above, we have developed a prototype of this integer linear programming system in R. We currently do not consider temporal locality between different accesses to the same variable.

%GABRIEL: otra pregunta sin relevancia para el artículo: cuánto tarda el sistema de programación lineal en analizar cada benchmark, aproximadamente?
\section{Experimental results}\label{sec:results}
\subsection{Configuration}
Besides implementing the architecture proposed in the simulator, it is important to test its performance and compare it with other architectural alternatives. For this purpose, we have chosen the PolyBench/C suite~\cite{polybench:source}. The main advantages of this suite are its open source license and having simple kernels to analyze for our model, i.e., nested loops with affine accesses, making it simple to calculate the number of reads and writes of each variable.

Another important aspect in order to perform a correct configuration of our architecture is the selection of the characteristics of our memory modules, i.e., energy consumption and latencies. For this purpose we have used CACTI~\cite{cacti} and NVSim~\cite{nvsim}. Actually, NVSim is a CACTI extension, since it works using this tool, but adding other features and technologies. These tools have been used to validate the configurations of our architectures.

\begin{table*}[htbp]
\small
\centering
\caption{Comparison of different technologies. Each row corresponds to a certain cache or scratchpad memory configuration.}

%[OK]GABRIEL: cambia un poco esta tabla para que no queden huecos raros entre filas
\begin{tabular}{|r|r|r|r|l|r|r|r|}
\hline
\multicolumn{8}{|c|}{\textbf{SRAM (caches)}} \\ \hline
\multicolumn{1}{|l|}{\textbf{cache}} & \multicolumn{1}{c|}{\textbf{mm$^2$}} & \multicolumn{3}{c|}{\textbf{Latencies (ns)}} & \multicolumn{3}{c|}{\textbf{Energies}}\\ \cline{ 1- 1}\cline{ 3- 8}
\multicolumn{1}{|c|}{\textbf{kB}} &  & \multicolumn{1}{c|}{\textbf{Read}} & \multicolumn{1}{c|}{\textbf{Miss}} & \multicolumn{1}{c|}{\textbf{Write}} & \multicolumn{1}{c|}{\textbf{Hit/miss (pJ)}} & \multicolumn{1}{c|}{\textbf{Write (pJ)}} & \multicolumn{1}{c|}{\textbf{Leak (mW)}} \\ \hline
256 & 0.229 & 2.258 & 0.083 & \multicolumn{1}{r|}{1.588} & 72 & 25 & 336.330 \\ \hline
512 & 0.380 & 2.669 & 0.107 & \multicolumn{1}{r|}{1.996} & 112 & 21 & 600.112 \\ \hline
1024 & 0.741 & 3.452 & 0.144 & \multicolumn{1}{r|}{2.773} & 214 & 36 & 1180.407 \\ \hline
2048 & 1.343 & 9.989 & 0.149 & \multicolumn{1}{r|}{7.941} & 378 & 24 & 2141.436 \\ \hline
4096 & 2.619 & 11.52 & 0.222 & \multicolumn{1}{r|}{9.037} & 383 & 290 & 4288.790 \\ \hline
\multicolumn{8}{|c|}{\textbf{STT-RAM (scratchpads)}} \\ \hline
\multicolumn{1}{|l|}{\textbf{SPM}} & \multicolumn{1}{l|}{\textbf{mm$^2$}} & \multicolumn{3}{c|}{\textbf{Latencies (ns)}} & \multicolumn{3}{c|}{\textbf{Energies}} \\ \cline{ 1- 1}\cline{ 3- 8}
\multicolumn{1}{|c|}{\textbf{kB}} & & \multicolumn{1}{c|}{\textbf{Read}} & \multicolumn{1}{c|}{\textbf{Miss}} & \multicolumn{1}{c|}{\textbf{Write}}
	%\hspace{-0.5cm}
%[OK]GABRIEL: quedaría mejor si la columna N/A fuera la central, para ser simétrica con respecto a la tabla SRAM. [OK]Quizá también quedaría mejor la tabla SRAM encima de la STTRAM.
 & \multicolumn{1}{c|}{\textbf{Read (pJ)}} & \multicolumn{1}{c|}{\textbf{Write (pJ)}} & \multicolumn{1}{c|}{\textbf{Leak (mW)}} \\ \hline
1024 & 0.183 & 2.221 & N.A. & 5.686 & 195.251 & 205.024 & 84.809 \\ \hline
2048 & 0.348 & 2.364 & N.A. & 5.744 & 228.512 & 242.614 & 146.194 \\ \hline
4096 & 0.696 & 2.499 & N.A. & 5.812 & 276.137 & 290.231 & 292.389 \\ \hline
8192 & 1.311 & 3.055 & N.A. & 6.038 & 388.324 & 383.871 & 568.592 \\ \hline
16384 & 2.488 & 5.036 & N.A. & 7.739 & 516.687 & 465.678 & 640.935\\ \toprule 
\end{tabular}
\label{tab:technologies}
\end{table*}

We have tested our proposal using two different architectures: a general purpose one, and a modified architecture including scratchpad memories. The general purpose architecture consists of an x86 processor and a traditional memory hierarchy: 
%[OK]GABRIEL: siempre son dos niveles en la configuración básica, no?
%different
two cache levels and a main memory (see Figure \ref{fig:archconvgem5}). Regarding the modified architecture (see Figure \ref{fig:archspmgem5}), the most remarkable detail of the configuration is the inclusion of a scratchpad memory between the CPU and main memory, as explained before. In order to perform meaningful comparisons, this scratchpad memory replaces the last-level cache memory in the traditional memory hierarchy. This is an area equivalent replacement, in other words, the L2 cache, is replaced with a scratchpad memory which occupies roughly the same die area. As a proof of concept of our system we explore, besides SRAM, the usage of STT-RAM memories, providing better energy consumption and capacity characteristics. Table \ref{tab:technologies} summarizes the different memory configurations used in our tests,
%[OK]GABRIEL: este párrafo creo que repite conceptos de los dos anteriores
% Table \ref{tab:technologies} shows the different configurations we have obtained for the memories. In order to compare fairly the different architectures, we have calculated the area of a cache memory with a certain capacity and looked for a STT-RAM scratchpad memory with the same size. In this manner, we have obtained five different configurations both cache and scratchpad memory.
%[OK]GABRIEL
%Regarding the power and energy of these configurations, Table \ref{tab:technologies}
including the energy consumed for both read and write accesses and also the leakage power of each module. Note the huge difference in the leakage power of both technologies. STT-RAM also has a higher density, increasing the size of the memory modules 
%[OK]Hence, the configuration of our experiment enables the use of higher capacity memories due to the shift of technology. besides the leakage power is improved 
by a factor of approximately four; however in detriment of the access latencies. The results of this experiment exemplify the energetic and temporal trade-off in the technological selection.

\subsection{Results}
Throughout this experiment we have talked about different features: area, latencies and energy, mostly. In the section above we talked about the different configurations of the memory hierarchy, also referring to the different technologies. Nevertheless, the theoretical premises may not be reflected in the execution of the benchmarks.

In order to measure the execution time of the programs, we set breakpoints at the interest regions, focusing on the kernel of the program (the SCoPs~\cite{bib:scops} in PolyBench codes). Thus gem5 gives us statistics about the execution time and energy consumed 
%in pJ 
by the memory hierarchy. In order to obtain homogeneous results, we have used gem5 to obtain the number of references to SPM, both reads and writes, as well as the execution time of the kernel of the program. With this data it is possible to calculate the dynamic energy consumed by scratchpad memories (see Equation \ref{eq:energyspm} below). 
%GABRIEL: qué diferencia hay entre los resultados de McPAT y los que obtendrías multiplicando el número de accesos a memoria por la energía por acceso y sumando la energía estática? O el problema es que no está claro cómo extraer el número de accesos a cache/memoria a partir de las estadísticas de gem5?
%MARCOS: básicamente es lo que comentas. En la comunidad de gem5 la gente se 'queja' de esto. De ahí que hiciese la solución "made na casa" para traducir de gem5 a McPAT.
Nevertheless, energetic results given by gem5 are quite scarce regarding memories, and in order to obtain more detailed results, we have used McPAT~\cite{bib:mcpat}.
%One issue is the lack of integration between McPAT and gem5. 
For this purpose we have developed a novel parser (gem5McPATparse~\cite{bib:gem5mcpat}) that searches the output files of gem5 for the parameters and statistics that serve as input for McPAT, and translates them generating the corresponding XML input file. These translations are based on~\cite{endo:tel-01285964}. In this way, the energy consumed by cache memories is calculated as the addition of the dynamic power and leakage power multiplied by the execution time (see Equation \ref{eq:energydyn}).

Finally, static energy is calculated using the architectural parameters provided by NVSim and the execution time of a particular program (see Equation \ref{eq:energystatic}). The addition of all equations is reflected in Equation \ref{eq:energytot}.

%GABRIEL: no veo claro el nombre de algunas de estas ecuaciones. P. ej., Edyn (imagino que de dynamic) incluye Pcache, pero Pcache es la potencia total, incluyendo energía estática. Deberían estar claramente diferenciados los componentes estático y dinámico de MM, cache y SPM.
%[MARCOS]: ahora? básicamente he intentado normalizar como comentas
\begin{equation}\label{eq:energystatic}
	\begin{aligned}
    	 E_{STATIC}(t) & = t_{exec} * P_{leakage}\\
    \end{aligned}
\end{equation}

\begin{equation}\label{eq:energyspm}
	\begin{aligned}
    	 E_{SPM} & = N_{r}*E_{SPM\_r} + N_{w}*E_{SPM\_w}\\
    \end{aligned}
\end{equation}

\begin{equation}\label{eq:energydyn}
	\begin{aligned}
    	P_{cache} & = P_{cache\_LEAK} + P_{cache\_DYN} \\
        E_{DYNcache} & = P_{cache}*t_{exec}\\
    	E_{DYN}(t) & = E_{MM} + E_{DYNcache} + E_{SPM}\\
    \end{aligned}
\end{equation}

\begin{equation}\label{eq:energytot}
	\begin{aligned}
    	 E_{TOT}(t) & =  E_{STATIC}(t) + E_{DYN}(t)\\
    \end{aligned}
\end{equation}

%[OK]GABRIEL: quita la cita al cluster y da todos los detalles de su hardware etc. Creo que están en mi paper de CGO 2016, si quieres copiarlos aquí.
Each execution was performed on an Intel Xeon E5-2660 Sandy Bridge 2.20 Ghz node, with 64 GB of RAM. All the results have been normalized with respect to the results obtained for \texttt{cache\_256kB}, which corresponds to the first row of Table~\ref{tab:technologies}. The most remarkable aspects observed in these executions are commented in the following paragraphs. 

\begin{figure}
	\includegraphics[scale=0.55]{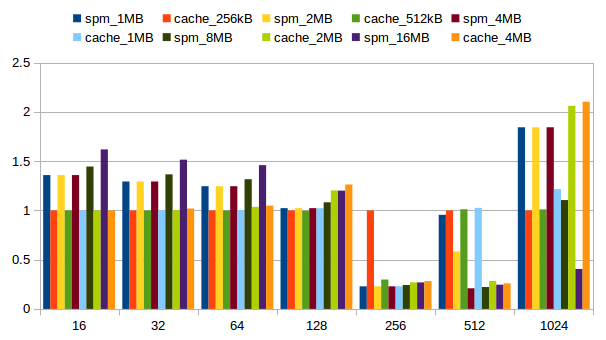}
    \caption{Normalized execution time of the \texttt{2mm} benchmark}
    \label{fig:time1}
\end{figure}

\begin{figure}
	\includegraphics[scale=0.55]{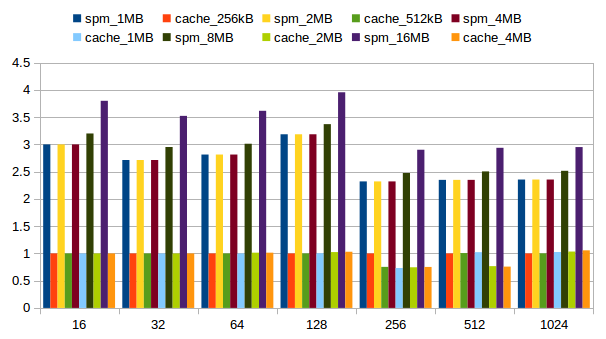}
    \caption{Normalized execution time of the \texttt{bicg} benchmark}
    \label{fig:time2}
\end{figure}

%[OK]GABRIEL: cambia esta enumeración por tres párrafos independientes. Empieza cada uno de ellos con variaciones de "Regarding performance... " o "The dynamic energy ...".
Regarding \textbf{temporal performance}, scratchpad architectures present worse results with small matrix sizes, in other words, with a small working set. The explanation for this issue is the fact that the decision system forces arrays with non cache-friendly accesses to be allocated to scratchpad memory. As a consequence, for small working sets where L1 cache could allocate this data without damaging locality the scratchpad memory increases latencies and dynamic energy consumed. The reason is that the SPM has been designed as a replacement for L2, but it bypasses L1 in these situations. 
%GABRIEL: no tengo claro cómo encaja esta frase en este párrafo. Creo que estamos hablando de algo diferente.
%Hence, in future refinements of this approach it would be convenient to include 1-dimensional arrays in the cache-friendly analysis.
%[MARCOS_old]: comento que estaría bien incluirlos porque en nuestro approach dijimos que los vectores unidimensinales NO los considerábamos cache friendly aunque lo fuesen, de modo que como a priori siempre van a caber en memoria SPM, el decisor los alojará ahí directamente
Nevertheless, when the size of these arrays increases, the decision system plays a major role since the L1 cache can no longer allocate all the desired data, provoking conflicts when accessing new elements. In these cases, using scratchpad memories mitigates cache memory penalties caused by non cache-friendly accesses and capacity issues. This can be observed in Figure~\ref{fig:time1} for the \texttt{2mm} benchmark. Still, there are also cases where this behavior is missing, as shown in Figure~\ref{fig:time2} for the \texttt{bicg} code.
%[DONE (ver párrafo anterior)]GABRIEL: no está explícito en ninguna parte cómo se realiza la normalización en las figuras, y no soy capaz de deducirlo. Sé que lo habíamos hablado pero no lo recuerdo.
%[OK]GABRIEL: no me queda demasiado claro por qué en este caso no funciona bien el SPM.
%MARCOS: mejor así un poco más verbose?
%MARCOS 2: ahora?
In this case, the decision system does not provide any performance enhancement since the data set works properly through cache memory, and losing the last-level cache is counterproductive.
% Analyzing the output of the decision system for all the benchmarks, we concluded that the lost of performance can be caused by: (i) not using SPM,
% or (ii) the disadvantageous use of SPM for certain variables. In the first case (i), if the system decides not to use the SPM, the performance will suffer due to the increase of accesses to main memory. This decision can be the product of the space left in the scratchpad or due to mark variables that would fit in the SPM as cache-friendly. On the second case (ii), some accesses can be harmful for performance if they fit in L1 cache, since the latencies increase. For instance, 1D vectors usually have a cache-friendly pattern, however we are not considering them for being allocated directly in main memory. This consideration was meant to decrease the number of conflicts in cache memories. Nevertheless results demonstrate that there is no benefit in allocating these kind of vectors in SPM.
%GABRIEL: mi problema es que ya hemos comentado lo de los arrays 1D, en este párrafo, no creo que aporte nada repetirlo. A ver así.
Analyzing the output of the decision system for the benchmarks whose performance worsens with working set size, we conclude that the reason for this is that some of the arrays which should be allocated to SPM from a locality point of view do not fit the available SPM space. Future versions of the decision system need to incorporate tiling transformations, or some means to store large arrays in a piecewise fashion into the SPM so that locality can be exploited in these cases.
    
%[OK]GABRIEL: no está nada claro cómo se relacionan las etiquetas de las figuras con las configuraciones de la tabla. Quizá una buena manera de clarificarlo sería cambiar las etiquetas de las figuras, donde pones "spm_cX" podrías usar "spm_256k" ... "spm_16M". Y lo mismo para "cache_cX". 

\begin{figure}
	\includegraphics[scale=0.55]{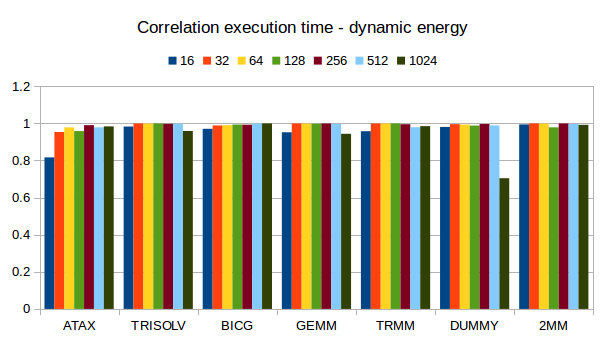}
    \caption{Correlation between execution time and dynamic energy for all benchmarks}
    \label{fig:corr}
\end{figure}

\begin{figure}[h]
	\includegraphics[scale=0.55]{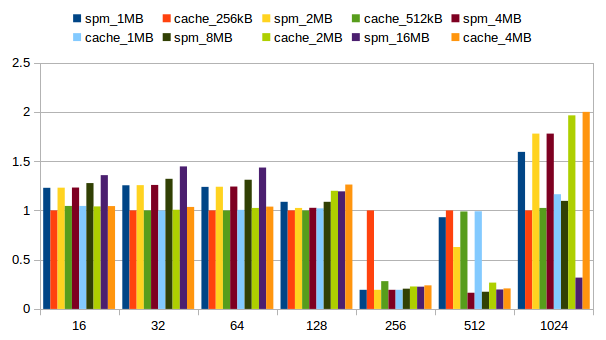}
    \caption{Normalized dynamic energy of the \texttt{2mm} benchmark}
    \label{fig:dyn1}
\end{figure}

\begin{figure}[h]
	\includegraphics[scale=0.55]{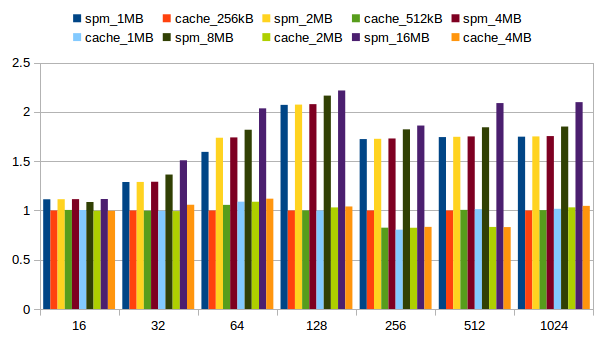}
    \caption{Normalized dynamic energy of the \texttt{bicg} benchmark}
    \label{fig:dyn2}
\end{figure}

Following a similar pattern to temporal performance, the \textbf{dynamic energy} in architectures with scratchpads is higher for small working sets, since the L1 cache has lower dynamic energy per access and no capacity issues arise. Similarly, when increasing the size of arrays, architectures without a scratchpad memory issue more accesses to main memory, raising the dynamic energy consumed (see Figures~\ref{fig:dyn1} and~\ref{fig:dyn2}). Nonetheless, it is also remarkable what we can observe in \texttt{bicg}: there is no performance improvement and the dynamic energy is also significantly higher.
%GABRIEL: no acabo de entender esta última frase.
%A good explanation is that the penalty of accessing main memory is both temporal and in terms of dynamic energy consumed.
%MARCOS: no sé qué te parece lo de poner la correlación...
%GABRIEL: me parece bien, pero me genera dudas para aquéllos benchmarks que no cumplen la tendencia. ¿Qué pasa con trisolv-32? (y en general con todos los tamaños de trisolv) ¿Dummy 512 y 1024? ¿2mm 128? ?atax 16 y 32?
The increase in dynamic energy is caused by the same issues identified when analyzing performance. In fact, the correlation between execution time and dynamic energy is extremely high in all cases, as illustrated in Figure~\ref{fig:corr}.

\begin{figure}
	\includegraphics[scale=0.55]{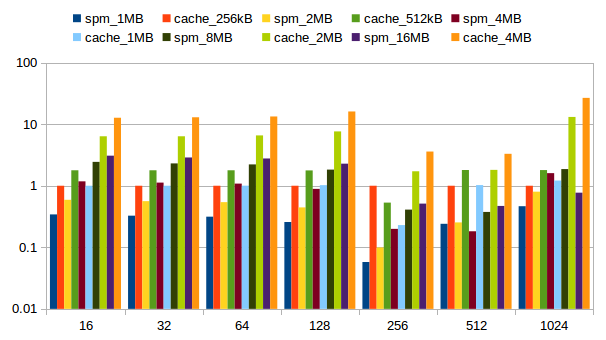}
    \caption{Normalized static energy of the \texttt{2mm} benchmark}
    \label{fig:static1}
\end{figure}

\begin{figure}
	\includegraphics[scale=0.55]{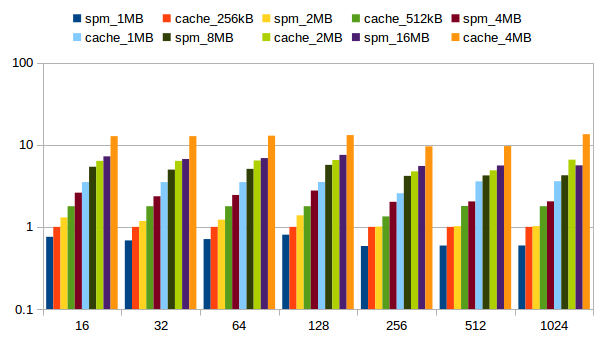}
    \caption{Normalized static energy of the \texttt{bicg} benchmark}
    \label{fig:static2}
\end{figure}

Regarding \textbf{static energy}, as expected, despite temporal differences, in general scratchpad memories have a lower static consumption due to the technology chosen (STT-RAM against SRAM). This situation can be observed in Figures \ref{fig:static1} and \ref{fig:static2}.
    
\begin{figure*}[h]
	\centering
    
   \includegraphics[scale=0.36]{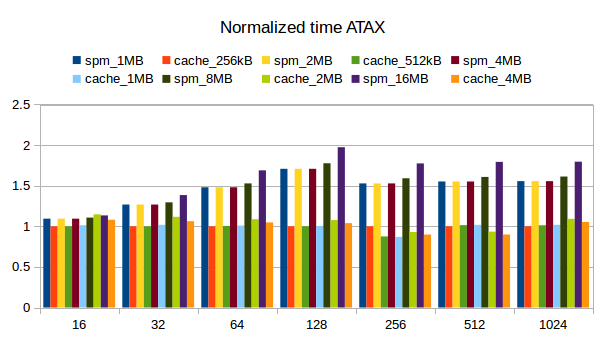}
   \includegraphics[scale=0.36]{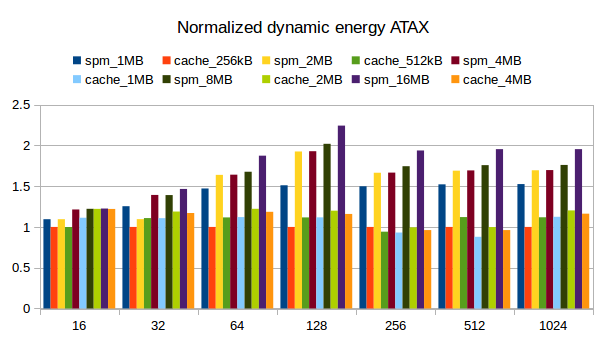}
   \includegraphics[scale=0.36]{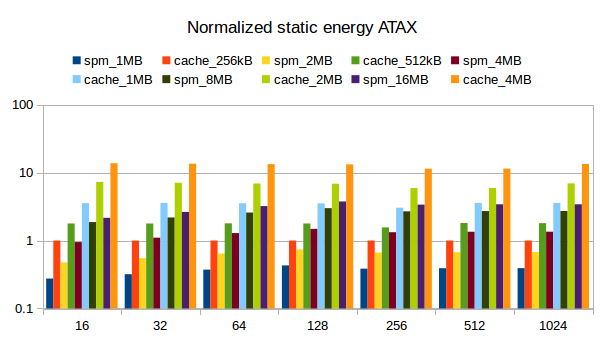}
   
   \includegraphics[scale=0.36]{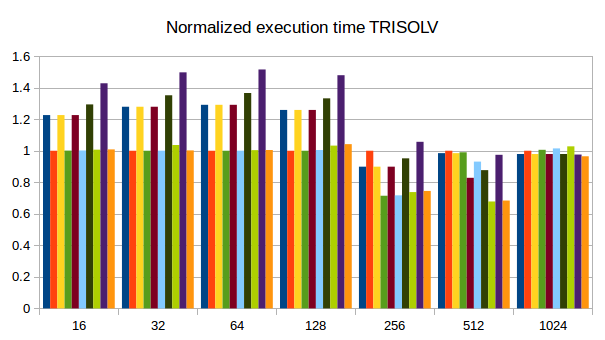}
   \includegraphics[scale=0.36]{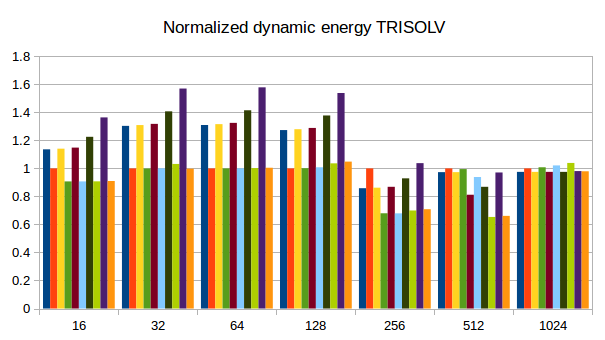}
   \includegraphics[scale=0.36]{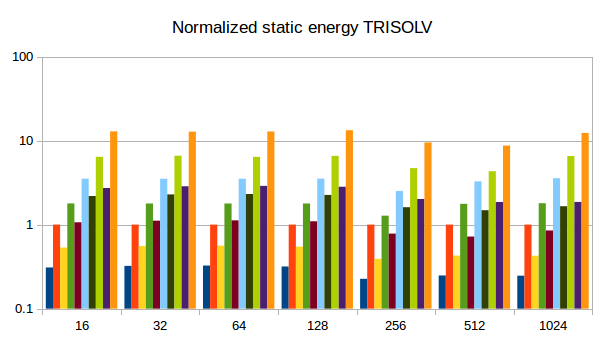} 
   
   \includegraphics[scale=0.36]{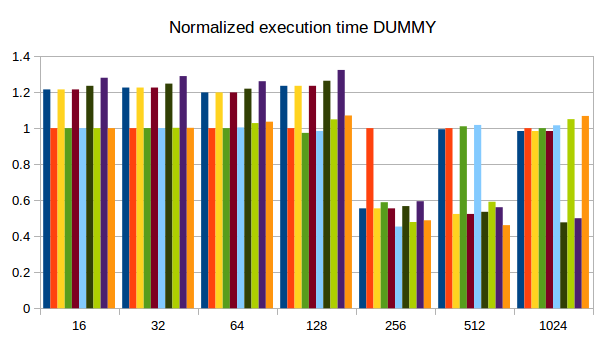}
   \includegraphics[scale=0.36]{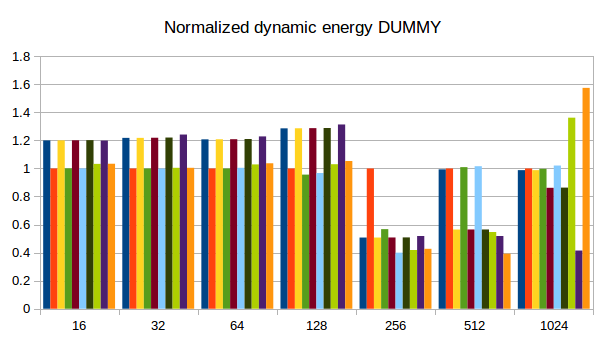}
   \includegraphics[scale=0.36]{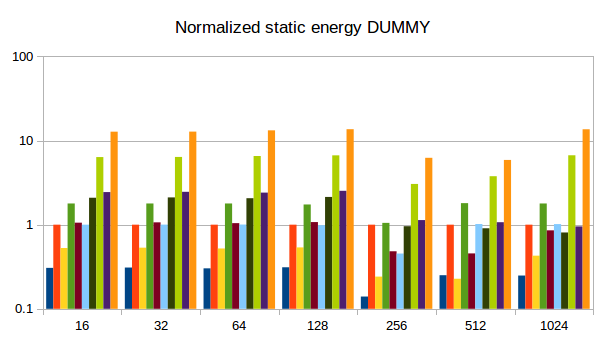}
   
  \includegraphics[scale=0.36]{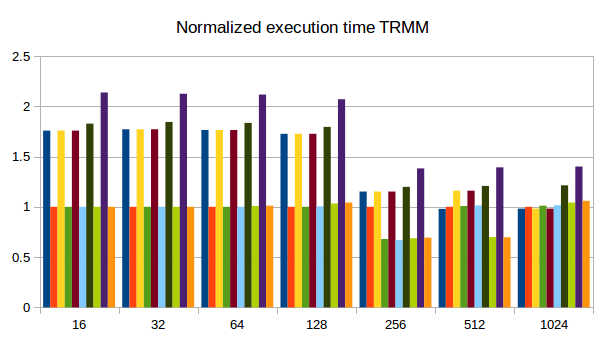}
  \includegraphics[scale=0.36]{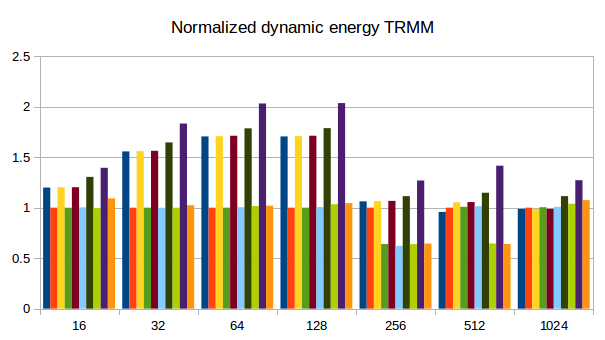}
  \includegraphics[scale=0.36]{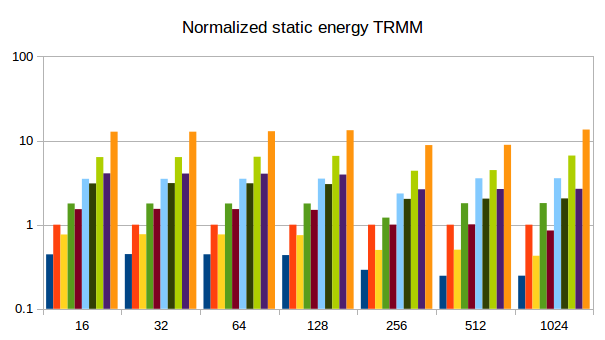} 
  
  \includegraphics[scale=0.36]{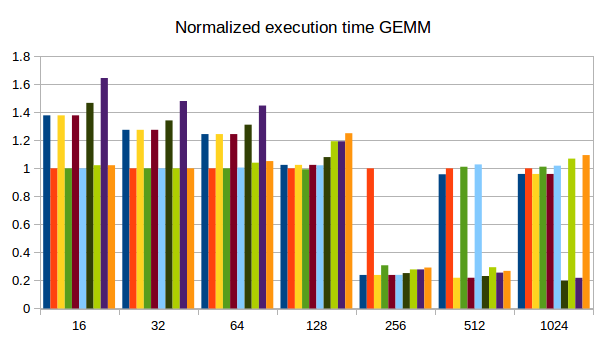}
  \includegraphics[scale=0.36]{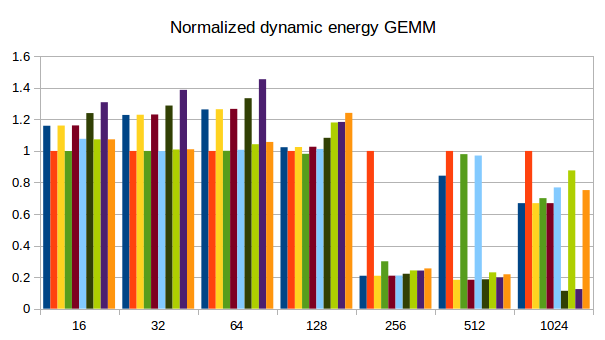}
  \includegraphics[scale=0.36]{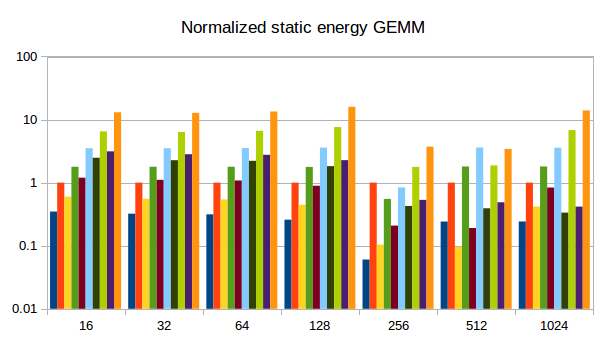}
  
  \caption{Results of all executions}
  \label{overall}
\end{figure*}

The executions for the remaining benchmarks are illustrated in Figure~\ref{overall}. It is readily observable the repetition of the patterns commented before in the rest of the benchmarks.
%GABRIEL: hay algún benchmark que no sea clasificable en ninguno de los dos patrones anteriores?
%MARCOS: no identifico otros patrones, ni veo anomalías extraordinarias a lo anteriormente comentado. De hecho, en las correlaciones que hacía

%%%%%%%%%%%%%%%%%%%%%%%%%%%%%%%%%
% TEMP
%[OK]GABRIEL: títulos en castellano

\section{Discussion}\label{sec:discussion}
The imperious requirement to use efficiently the available resources in our system presents interesting challenges. In this article general purpose architectures using different memory hierarchies are analyzed. For this purpose well-known modeling and simulation tools have been used in order to demonstrate the reliability of the results. A linear programming system to decide to which available module  a certain variable of the program must be allocated in order to improve energy consumption has been developed. We have compared cache and RAM memories equivalent in terms of area, using SRAM and STT-RAM technologies, respectively. Experimental results demonstrate that: (i) without a remarkable penalty in time, energy efficiency can be improved, mostly due to low leakage power in scratchpad memories using STT-RAM; (ii) the use of scratchpad memories with a simple decision algorithm can also improve temporal performance for certain benchmarks; (iii) there are some cases where the algorithm does not choose properly the location of variables, but this behavior is easily correctable considering the size of the L1 cache; and (iv) as the size of the working set grows, the energy improvements are more evident due to, in some cases, the success of the decider choosing the most profitable location for a variable, but mostly due to the weighing of leakage power.

\section{Acknowledgements}
This work is supported by the Ministry of Economy and Competitiveness of Spain and FEDER funds of the EU (Project TIN2013-42148-P).


%\bibliographystyle{IEEEtran}
\bibliographystyle{IEEEtran}
\bibliography{main.bib} 

\begin{thebibliography}{10}
\providecommand{\url}[1]{#1}
\csname url@samestyle\endcsname
\providecommand{\newblock}{\relax}
\providecommand{\bibinfo}[2]{#2}
\providecommand{\BIBentrySTDinterwordspacing}{\spaceskip=0pt\relax}
\providecommand{\BIBentryALTinterwordstretchfactor}{4}
\providecommand{\BIBentryALTinterwordspacing}{\spaceskip=\fontdimen2\font plus
\BIBentryALTinterwordstretchfactor\fontdimen3\font minus
  \fontdimen4\font\relax}
\providecommand{\BIBforeignlanguage}[2]{{%
\expandafter\ifx\csname l@#1\endcsname\relax
\typeout{** WARNING: IEEEtran.bst: No hyphenation pattern has been}%
\typeout{** loaded for the language `#1'. Using the pattern for}%
\typeout{** the default language instead.}%
\else
\language=\csname l@#1\endcsname
\fi
#2}}
\providecommand{\BIBdecl}{\relax}
\BIBdecl

\bibitem{bib:dennard}
\BIBentryALTinterwordspacing
H.~Esmaeilzadeh, E.~Blem, R.~St.~Amant, K.~Sankaralingam, and D.~Burger,
  ``{Dark Silicon and the End of Multicore Scaling},'' in \emph{Proceedings of
  the 38th Annual International Symposium on Computer Architecture}, New York,
  NY, USA, 2011, pp. 365--376. [Online]. Available:
  \url{http://doi.acm.org/10.1145/2000064.2000108}
\BIBentrySTDinterwordspacing

\bibitem{bib:yan}
L.~Yan, L.~Dongsheng, Z.~Duoli, D.~Gaoming, W.~Jian, G.~Minglun, W.~Haihua, and
  G.~Luofeng, ``{Performance evaluation of the memory hierarchy design on CMP
  prototype using FPGA},'' in \emph{IEEE 8th International Conference on ASIC},
  Oct 2009, pp. 813--816.

\bibitem{spmvscache}
{Vivy Suhendra, Chandrashekar Raghavan, Tulika Mitra}, ``{Integrated Scratchpad
  Memory Optimization and Task Scheduling for MPSoC Architecture},''
  \emph{School of computing. National University of Singapore}, 2006.

\bibitem{spmvscache2}
B.~Anuradha and C.~Vivekanandan, ``Usage of scratchpad memory in embedded
  systems 2014; state of art,'' in \emph{Third International Conference on
  Computing Communication Networking Technologies (ICCCNT)}, July 2012, pp.
  1--5.

\bibitem{bib:cell}
M.~Gschwind, H.~P. Hofstee, B.~Flachs, M.~Hopkins, Y.~Watanabe, and
  T.~Yamazaki, ``Synergistic processing in cell's multicore architecture,''
  \emph{IEEE Micro}, vol.~26, no.~2, pp. 10--24, March 2006.

\bibitem{bsc}
{Barcelona Supercomputing Center (BSC)}, ``{Cell Superscalar (CellSs) User’s
  Manual},'' \url{http://www.bsc.es/media/2296.pdf}, 2009.

\bibitem{bib:cellexamples}
M.~Takayama and R.~Sakai, ``{Parallelization Strategy for CELL TV},'' in
  \emph{1st Workshop on Applications for Multi and Many Core Processors}, 2010.

\bibitem{bib:ps2}
{T. M. Conte}, \emph{{Computer Architecture: A Quantitative Approach. Appendix
  E: Embedded Systems}}, 5th~ed.\hskip 1em plus 0.5em minus 0.4em\relax San
  Francisco, CA, USA: Morgan Kaufmann Publishers Inc., 2011.

\bibitem{bib:dsp}
Y.~Lin, H.~Lee, M.~Woh, Y.~Harel, S.~Mahlke, T.~Mudge, C.~Chakrabarti, and
  K.~Flautner, ``Soda: A low-power architecture for software radio,''
  \emph{SIGARCH Comput. Archit. News}, vol.~34, no.~2, pp. 89--101, May 2006.

\bibitem{bib:knights}
A.~Sodani, ``{Intel® Xeon Phi™ Processor "Knights Landing" Architectural
  Overview},''
  \url{https://www.nersc.gov/assets/Uploads/KNL-ISC-2015-Workshop-Keynote.pdf},
  2015.

\bibitem{ibm}
{IBM}, ``{IBM 6X86MX Microprocessor},''
  \url{http://datasheets.chipdb.org/IBM/x86/6x86MX/mx_full.pdf}, 1998.

\bibitem{gem5accuracy}
A.~Butko, R.~Garibotti, L.~Ost, and G.~Sassatelli, ``Accuracy evaluation of
  gem5 simulator system,'' in \emph{7th International Workshop on
  Reconfigurable Communication-centric Systems-on-Chip (ReCoSoC)}, July 2012,
  pp. 1--7.

\bibitem{gem5:errors}
A.~Gutierrez, J.~Pusdesris, R.~Dreslinski, T.~Mudge, C.~Sudanthi, C.~Emmons,
  M.~Hayenga, and N.~Paver, ``Sources of error in full-system simulation,'' in
  \emph{Proceedings of the IEEE International Symposium on Performance Analysis
  of Systems and Software (ISPASS)}, March 2014, pp. 13--22.

\bibitem{bib:cacheenergy}
{R. Banakar, S. Steinke, L. Bo-Sik, M. Balakrishnan, and P. Marwedel},
  ``Scratchpad memory: Design alternative for cache on-chip memory in embedded
  systems.'' \emph{In Proceedings of the 10th International Symposium on
  Hardware/Software Codesign}, pp. 73--78, 2002.

\bibitem{gabri}
\BIBentryALTinterwordspacing
G.~Rodr\'{\i}guez, J.~Touri\~{n}o, and M.~T. Kandemir, ``{Volatile STT-RAM
  Scratchpad Design and Data Allocation for Low Energy},'' \emph{ACM Trans.
  Archit. Code Optim.}, vol.~11, no.~4, pp. 38:1--38:26, Dec. 2014. [Online].
  Available: \url{http://doi.acm.org/10.1145/2669556}
\BIBentrySTDinterwordspacing

\bibitem{polybench:source}
L.~Pouchet, ``{PolyBench/C: the Polyhedral Benchmark suite},''
  \url{http://web.cse.ohio-state.edu/~pouchet/software/polybench/}, 2015.

\bibitem{cacti}
S.~J.~E. Wilton and N.~P. Jouppi, ``{CACTI: an enhanced cache access and cycle
  time model},'' \emph{IEEE Journal of Solid-State Circuits}, vol.~31, no.~5,
  pp. 677--688, May 1996.

\bibitem{nvsim}
X.~Dong, C.~Xu, Y.~Xie, and N.~P. Jouppi, ``{NVSim: A Circuit-Level
  Performance, Energy, and Area Model for Emerging Nonvolatile Memory},''
  \emph{IEEE Transactions on Computer-Aided Design of Integrated Circuits and
  Systems}, vol.~31, no.~7, pp. 994--1007, July 2012.

\bibitem{bib:scops}
A.~Kumar and S.~Pop, ``{SCoP Detection: A Fast Algorithm for Industrial
  Compilers},'' in \emph{6th International Workshop on Polyhedral Compilation
  Techniques on IMPACT}, January 2016.

\bibitem{bib:mcpat}
S.~Li, J.~H. Ahn, R.~D. Strong, J.~B. Brockman, D.~M. Tullsen, and N.~P.
  Jouppi, ``{McPAT: An Integrated Power, Area, and Timing Modeling Framework
  for Multicore and Manycore Architectures},'' in \emph{Proceedings of the 42nd
  Annual IEEE/ACM International Symposium on Microarchitecture (MICRO 42)},
  2009, pp. 469--480.

\bibitem{bib:gem5mcpat}
M.~Horro, ``{gem5McPATparse},''
  \url{https://github.com/markoshorro/gem5McPATparse}, 2016.

\bibitem{endo:tel-01285964}
\BIBentryALTinterwordspacing
F.~Endo, ``{Online Auto-Tuning for Performance and Energy through
  Micro-Architecture Dependent Code Generation},'' Theses, {Universit{\'e}
  Grenoble Alpes}, Sep. 2015. [Online]. Available:
  \url{https://tel.archives-ouvertes.fr/tel-01285964}
\BIBentrySTDinterwordspacing

\end{thebibliography}

\end{document}